\documentclass[aps, prx, reprint, superscriptaddress, longbibliography]{revtex4-1}
\usepackage[dvipdfmx]{graphicx}
\usepackage{amsmath}
\usepackage{amssymb}
\usepackage{dcolumn}
\usepackage{color}
\usepackage{bm}
\usepackage{braket}
\usepackage{txfonts}

\newcommand{\up}{\uparrow}
\newcommand{\down}{\downarrow}

\begin{document}
\title{Spontaneous symmetry breaking in non-steady modes of open quantum many-body systems}
\author{Taiki Haga}
\email[]{taiki.haga@omu.ac.jp}
\affiliation{Department of Physics and Electronics, Osaka Metropolitan University, Sakai-shi, Osaka 599-8531, Japan}
\date{\today}

\begin{abstract}
In a quantum many-body system coupled to the environment, its steady state can exhibit spontaneous symmetry breaking when a control parameter exceeds a critical value.
In this study, we consider spontaneous symmetry breaking in non-steady modes of an open quantum many-body system.
Assuming that the time evolution of the density matrix of the system is described by a Markovian master equation, the dynamics of the system is fully characterized by the eigenmodes and spectrum of the corresponding time evolution superoperator.
Among the non-steady eigenmodes with finite lifetimes, we focus on the eigenmodes with the highest frequency, which we call the most coherent mode.
For a dissipative spin model, it is shown that the most coherent mode exhibits a transition from a disordered phase to a symmetry-broken ordered phase, even if the steady state does not show singular behavior.
We further argue that the phase transition of the most coherent mode induces a qualitative change in the decoherence dynamics of highly entangled states, i.e., the Schr\"odinger's cat states.
\end{abstract}

\maketitle

\section{Introduction}

Recent advances in quantum engineering have made it possible to precisely control atomic, molecular, and optical systems, which include ultracold atoms in optical lattices \cite{Bloch-08-1, Bloch-08-2, Syassen-08, Bloch-12, Ritsch-13}, trapped ions \cite{Lanyon-09, Barreiro-11, Blatt-12}, Rydberg atoms \cite{Low-12, Browaeys-20, Schauss-12, Sanchez-18, Lienhard-18, Borish-20}, and coupled optical cavities \cite{Hartmann-06, Baumann-10, Eichler-14, Rodriguez-16}.
Decoherence due to coupling with the environment is inevitable in these systems.
The nonequilibrium dynamics of open quantum many-body systems is highly complex and largely unexplored due to the intricate interplay of coherent Hamiltonian dynamics and dissipative dynamics arising from interactions with the environment.

Spontaneous symmetry breaking (SSB) is a pivotal concept in condensed matter physics and statistical mechanics.
The ground state of some quantum many-body systems, such as the transverse field Ising model and the Bose-Hubbard model, is known to exhibit a phase transition from a disordered phase to a symmetry-broken ordered phase \cite{Sachdev}.
In a typical open quantum many-body system, a unique steady state is realized in the long-time limit, so it is natural to consider SSB in the steady state.
Such a steady state is not necessarily at thermal equilibrium but can be a nonequilibrium state maintained by a balance between external driving and energy dissipation to the environment.
Phase transitions in nonequilibrium steady states of open quantum many-body systems are known as dissipative phase transitions and have attracted much attention in recent years \cite{Tomadin-11, Lee-11, Kessler-12, Honing-12, Horstmann-13, Torre-13, Lee-13, Ludwig-13, Carr-13, Sieberer-14, Marcuzzi-14, Weimer-15, Maghrebi-16, Sieberer-16, Biondi-17, Rota-18, Minganti-18, Ferreira-19, Young-20, Domokos-17, Fitzpatrick-17, Imamoglu-18}.

The dynamics of Markovian open quantum systems is governed by the Gorini--Kossakowski--Sudarshan--Lindblad (GKSL) master equation \cite{Lindblad-76, Gorini-76}.
The superoperator that generates the time evolution of a density matrix is called Liouvillian.
The dynamics of an open quantum system is completely determined by the eigenmodes and spectrum of the Liouvillian.
The steady states correspond to eigenmodes with zero eigenvalues.
Therefore, the dissipative phase transition is interpreted as a transition of a Liouvillian eigenmode.

In this study, we consider the phase structure of non-steady eigenmodes that have nonzero decay rates.
These eigenmodes are irrelevant to the long-time behavior of the system but contribute to the transient dynamics toward a steady state.
We here ask what happens when non-steady eigenmodes exhibit SSB as certain control parameters are varied.
Phase transitions of non-steady eigenmodes can significantly alter the relaxation dynamics of the system.
However, even if some non-steady eigenmodes undergo a phase transition, the steady state does not necessarily exhibit singular behavior.
Phase transitions of non-steady eigenmodes can provide a new mechanism for dynamical phase transitions in open quantum many-body systems.

The purpose of this study is to demonstrate that non-steady eigenmodes of a well-studied open quantum many-body system exhibit previously unrecognized SSB.
In particular, we consider one of the simplest models of open quantum many-body systems, the transverse field Ising model under dephasing.
The steady state of this system is an infinite-temperature state, regardless of the strength of the transverse field or dephasing.
Among the non-steady eigenmodes, we focus on the one with the highest frequency and call it the most coherent mode of the Liouvillian.
Mean-field analysis for the infinite-dimensional case and finite-size numerics for the one-dimensional case show that the most coherent mode exhibits a phase transition from a disordered (paramagnetic) phase to a symmetry-broken (ferromagnetic) phase at a critical transverse field that depends on the dephasing rate.
This SSB affects the relaxation dynamics of highly entangled states.
Suppose that an initial state is taken to be an equal superposition of two symmetry-broken states, which is known as the ``Schr\"odinger's cat state".
As the disorder-to-order phase transition of the most coherent mode occurs, the early dynamics of the density matrix shows a crossover from strongly-damped relaxation to under-damped relaxation with temporal oscillations.
This study reveals a typical situation in which SSB of non-steady eigenmodes causes a qualitative change in the decoherence dynamics of an open quantum system.

This paper is organized as follows.
In Sec.~\ref{sec:transition_non_steady_eigenmodes}, we review the quantum master equation for open quantum systems and define phase transitions in non-steady eigenmodes.
In Sec.~\ref{sec:model}, we introduce the transverse field Ising model under dephasing.
The structures of eigenmodes and spectra are discussed for a small system.
Section \ref{sec:mean_field} presents the results of the mean-field analysis for the infinite-dimensional case.
The phase diagram for the transverse field and dephasing rate is determined.
In Sec.~\ref{sec:numerical_1D}, the critical field in the one-dimensional case is determined by numerically solving the quantum master equation, and qualitative agreement with the mean-field calculation is confirmed.
In Sec.~\ref{sec:relaxation}, we discuss the effect of the transition of the most coherent mode on the relaxation dynamics of a highly entangled cat state.
Section \ref{sec:conclusion} is devoted to discussion and conclusions.

\section{Phase transition of non-steady modes}
\label{sec:transition_non_steady_eigenmodes}

The time evolution of a Markovian open quantum system is described by the GKSL quantum mater equation \cite{Lindblad-76, Gorini-76}:
\begin{equation}
	i\frac{d \rho}{dt} = \mathcal{L}(\rho) := [H, \rho] + i \sum_{\nu} \left( L_{\nu} \rho L_{\nu}^{\dag} - \frac{1}{2} \{ L_{\nu}^{\dag}L_{\nu}, \rho \} \right),
	\label{master_eq_1}
\end{equation}
where $\rho$ is the density matrix of the system, $[A,B] = AB-BA$, $\{A,B\} = AB+BA$, and $L_{\nu}$ are called jump operators.
The index $\nu$ of $L_{\nu}$ represents the types of dissipation or the lattice sites.
The superoperator $\mathcal{L}$ that generates the time evolution of $\rho$ is known as the Liouvillian.
The quantum master equation \eqref{master_eq_1} is justified when the time scale of the dynamics caused by the interaction with the environment is much longer than the time scale of the environment \cite{Breuer, Rivas}.
Here, we define the inner product between operators $A$ and $B$ by
\begin{equation}
	(A|B)  := \mathrm{Tr}[A^{\dag}B].
	\label{inner_product_operators}
\end{equation}
The adjoint operator of $\mathcal{L}$ is defined by $(A|\mathcal{L}(B))=(\mathcal{L}^\dag(A)|B)$, and given by
\begin{equation}
	\mathcal{L}^{\dag}(A) = [H, A] - i \sum_{\nu} \left( L_{\nu}^{\dag} A L_{\nu} - \frac{1}{2} \{ L_{\nu}^{\dag}L_{\nu}, A \} \right).
	\label{Liouvillian_dag}
\end{equation}

The right eigenmodes $\Phi_{\alpha}^{\mathrm{R}}$ and left eigenmodes $\Phi_{\alpha}^{\mathrm{L}}$ of $\mathcal{L}$ are defined by
\begin{equation}
	\mathcal{L}(\Phi_{\alpha}^{\mathrm{R}}) = \lambda_{\alpha} \Phi_{\alpha}^{\mathrm{R}}, \quad \mathcal{L}^{\dag}(\Phi_{\alpha}^{\mathrm{L}}) = \lambda_{\alpha}^* \Phi_{\alpha}^{\mathrm{L}},
	\label{right_left_eigenmode_eq}
\end{equation}
where $\lambda_{\alpha}$ denotes the $\alpha$th eigenvalue, and $\alpha=0, 1, ... , D^2-1$, provided that $D$ is the dimension of $\mathcal{H}$.
We assume that the eigenmodes are normalized as
\begin{equation}
	(\Phi_{\alpha}^{\mathrm{R}} | \Phi_{\alpha}^{\mathrm{R}})=1, \quad (\Phi_{\alpha}^{\mathrm{L}} | \Phi_{\beta}^{\mathrm{R}})=\delta_{\alpha \beta},
\end{equation}
for all $\alpha$ and $\beta$.
Note that, in general, the right eigenmodes $\Phi_{\alpha}^{\mathrm{R}}$ are not orthogonal to each other because $\mathcal{L}$ is not Hermitian.
The right eigenmodes $\Phi_\alpha^\mathrm{R}$ and eigenvalues $\lambda_{\alpha}$ have the following properties \cite{Minganti-18}:
\begin{enumerate}
	\item There is at least one right eigenmode $\Phi_0^\mathrm{R}$ with zero eigenvalue, $\mathcal{L}(\Phi_0^\mathrm{R})=0$, corresponding to the steady state.
	Here, $\Phi_0^\mathrm{R}$ is Hermitian and positive-semidefinite.
	\item All eigenvalues have non-positive imaginary parts, which guarantees the convergence of the density matrix to the steady state in the long-time limit.
	We assume that the eigenmodes are sorted such that $0 = |\mathrm{Im}[\lambda_0]| \leq |\mathrm{Im}[\lambda_1]| \leq \cdots \leq |\mathrm{Im}[\lambda_{D^2-1}]|$.
	\item $\mathrm{Tr}[\Phi_{\alpha}^{\mathrm{R}}]=0$ if $\lambda_\alpha \neq 0$, which follows from $\mathrm{Tr}[\mathcal{L}(\rho)]=0$.
	In general, $\Phi_{\alpha}^{\mathrm{R}}$ with nonzero $\lambda_\alpha$ is neither Hermitian nor positive-semidefinite.
	\item If $\mathcal{L}(\Phi_\alpha^\mathrm{R})=\lambda_\alpha \Phi_\alpha^\mathrm{R}$, then $\mathcal{L}((\Phi_\alpha^\mathrm{R})^\dag)=\lambda_\alpha^* (\Phi_\alpha^\mathrm{R})^\dag$.
	This means that the Liouvillian spectrum on the complex plane is symmetric with respect to the real axis.
\end{enumerate}

In the following, we see that the Liouvillian $\mathcal{L}$ can be interpreted as a non-Hermitian operator on an extended Hilbert space \cite{Prosen-12, Znidaric-15}.
Let $\{ \ket{i} \}_{i=1,...,D}$ be an orthonormal basis set of the Hilbert space $\mathcal{H}$ of the system.
An arbitrary operator $A$ can be mapped to a vector in $\mathcal{H} \otimes \mathcal{H}$ by
\begin{equation}
	A = \sum_{i,j=1}^D A_{ij} \ket{i} \bra{j} \ \to \ |A) = \sum_{i,j=1}^D A_{ij} \ket{i} \otimes \ket{j} \in \mathcal{H} \otimes \mathcal{H},
\end{equation}
where $A_{ij}=\braket{i|A|j}$.
We here denote a vector in $\mathcal{H} \otimes \mathcal{H}$ by the round ket $|...)$.
The inner product between $|A)$ and $|B)$ reads $(A|B)=\sum_{i,j=1}^D A_{ij}^* B_{ij}$ from Eq.~\eqref{inner_product_operators}.
In terms of this vector representation, the Liouvillian is rewritten as
\begin{equation}
	\mathcal{L} = H \otimes I - I \otimes H^{\mathrm{T}} + i \sum_{\nu} \mathcal{D}[L_\nu],
	\label{Liouvillian_ladder}
\end{equation}
with
\begin{equation}
	\mathcal{D}[L_\nu] = L_{\nu} \otimes L_{\nu}^* - \frac{1}{2} L_{\nu}^{\dag}L_{\nu} \otimes I - \frac{1}{2} I \otimes L_{\nu}^{\mathrm{T}} L_{\nu}^*,
\end{equation}
where $I$ represents the identity operator on $\mathcal{H}$, and $L_{\nu}^*$ is defined as $\braket{i|L_{\nu}^*|j} = \braket{i|L_{\nu}|j}^*$.
For simplicity of notation, the same notation $\mathcal{L}$ is used for the Liouvillian in the vector representation.
The master equation \eqref{master_eq_1} is then rewritten as the Schr\"odinger equation-like form,
\begin{equation}
	i\frac{d}{dt}|\rho) = \mathcal{L}|\rho).
	\label{master_eq_2}
\end{equation}
The right and left eigenmodes are written as $|\Phi_{\alpha}^{\mathrm{R}})$ and $|\Phi_{\alpha}^{\mathrm{L}})$ in the vector representation, respectively.
If the set of eigenmodes forms a basis of $\mathcal{H} \otimes \mathcal{H}$, the spectral decomposition of $\mathcal{L}$ reads
\begin{equation}
	\mathcal{L} = \sum_{\alpha} \lambda_{\alpha} | \Phi_{\alpha}^{\mathrm{R}}) (\Phi_{\alpha}^{\mathrm{L}} |.
	\label{L_spectral_decomposition}
\end{equation}
The time evolution of the density matrix $|\rho_t)$ is given by
\begin{equation}
	|\rho_t) = e^{-i\mathcal{L}t}|\rho_{\mathrm{ini}}) = \sum_{\alpha=0}^{D^2-1} c_{\alpha} e^{-i\lambda_{\alpha} t} |\Phi_{\alpha}^{\mathrm{R}}), \quad c_{\alpha} = (\Phi_{\alpha}^{\mathrm{L}}|\rho_{\mathrm{ini}}),
	\label{rho_expansion}
\end{equation}
where $|\rho_{\mathrm{ini}})$ is an initial state.

While the structure of the steady state $|\Phi_0^{\mathrm{R}})$ has been extensively studied in many previous studies, here we focus on the behavior of non-steady eigenmodes $|\Phi_{\alpha}^{\mathrm{R}})$ ($\mathrm{Im}[\lambda_\alpha]<0$).
Suppose that the Liouvillian $\mathcal{L}$ contains a control parameter $g$.
Let $|\Phi_{\alpha}^{\mathrm{R}}(g))$ and $|\Phi_{\alpha}^{\mathrm{L}}(g))$ be some (non-degenerate) right and left eigenmodes with a nonzero eigenvalue $\lambda_{\alpha}(g)$.
For a finite system, when $g$ is varied continuously, $|\Phi_{\alpha}^{\mathrm{R}}(g))$, $|\Phi_{\alpha}^{\mathrm{L}}(g))$, and $\lambda_{\alpha}(g)$ should also vary continuously almost everywhere in the parameter space.
It should be noted that in non-Hermitian systems, eigenvectors and eigenvalues can exhibit singular behavior when two eigenvalues collide at a certain parameter value, called the exceptional point \cite{Bender-98, Guo-09, Ruter-10, Heiss-12, El-Ganainy-18, Minganti-19, Ozdemir-19, Miri-19, Ashida-20}.
We assume that the eigenvalue $\lambda_{\alpha}(g)$ under consideration does not collide with another eigenvalue when $g$ is varied.
Suppose that for an operator $\mathcal{O}$ on $\mathcal{H} \otimes \mathcal{H}$, the following property is satisfied:
\begin{equation}
	\lim_{L \to \infty} ( \Phi_{\alpha}^{\mathrm{L}}(g)| \mathcal{O} | \Phi_{\alpha}^{\mathrm{R}}(g)) = 
	\begin{cases}
		M(g) \neq 0 & (g < g_\mathrm{c}); \\
		0 & (g \geq g_\mathrm{c}),
	\end{cases}
	\label{def_eigenmode_phase_transition}
\end{equation}
where $L$ represents the system size and $M(g)$ is a complex-valued analytic function.
Equation \eqref{def_eigenmode_phase_transition} defines the phase transition of a non-steady eigenmode.
The critical parameter $g_\mathrm{c}$ depends on the eigenmode under consideration.
Note that the phase transition of a non-steady eigenmode is not necessarily accompanied by a phase transition in the steady state.
In general, the phase transitions of non-steady eigenmodes do not affect the long-time behavior of the system, but they can affect the transient dynamics to the steady state.

\begin{figure}
	\centering
	\includegraphics[width=0.45\textwidth]{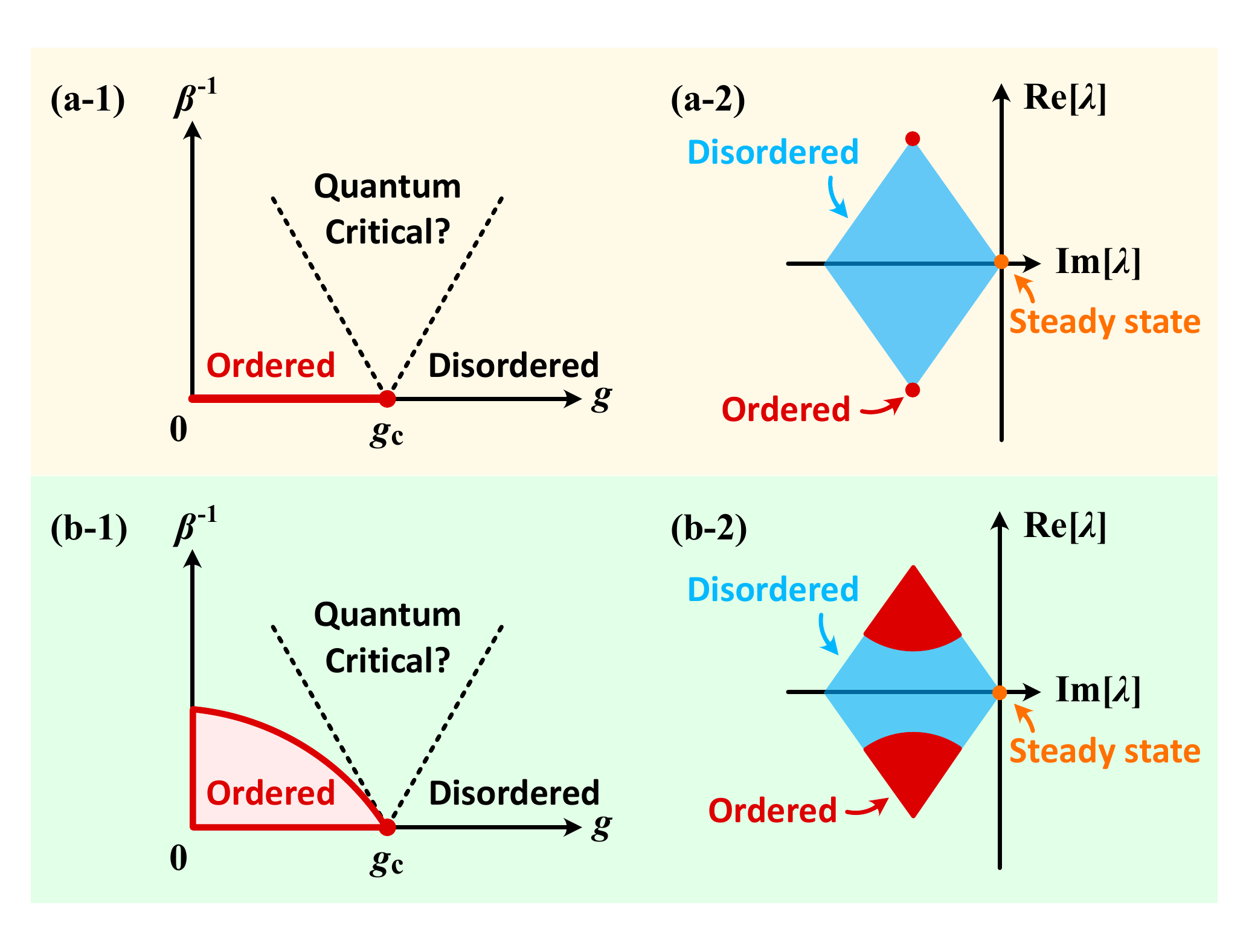}
	\caption{Schematic phase diagrams and Liouvillian spectra.
		(a) Case that the order exists only at zero temperature.
		(b) Case that the order exists at finite temperature.
		Figures (a-1) and (b-1) show the expected phase diagrams with respect to the control parameter $g$ and the fictitious temperature $\beta^{-1}$.
		The dashed lines represent the boundaries of the quantum critical region.
		Figures (a-2) and (b-2) show the Liouvillian spectra for $g < g_\mathrm{c}$.
		The red and blue regions represent ordered and disordered eigenmodes, respectively.}
	\label{Fig-QPT-diagram}
\end{figure}

Unfortunately, it is difficult to theoretically investigate the large-scale structure of each non-steady eigenmode.
Instead, we consider a quantity obtained by averaging with appropriate weights with respect to the eigenmodes.
By interpreting $\mathcal{L}$ as a non-Hermitian Hamiltonian on $\mathcal{H} \otimes \mathcal{H}$, we define the Liouvillian canonical average by
\begin{equation}
	\langle \mathcal{O} \rangle_\beta := \frac{\mathrm{Tr}\left[\mathcal{O} e^{-\beta \mathcal{L}}\right]}{\mathrm{Tr}\left[e^{-\beta \mathcal{L}}\right]} = \frac{\sum_\alpha e^{-\beta \lambda_\alpha} ( \Phi_{\alpha}^{\mathrm{L}}| \mathcal{O} | \Phi_{\alpha}^{\mathrm{R}})}{\sum_\alpha e^{-\beta \lambda_\alpha}},
	\label{def_canonical_average}
\end{equation}
where $\beta$ is a fictitious inverse temperature.
The canonical average defined by Eq.~\eqref{def_canonical_average} can be calculated by various theoretical methods developed in statistical mechanics and field theory, such as mean-field approximation, diagrammatic techniques, and renormalization group analysis.
Note that the expectation value $\langle \mathcal{O} \rangle_\beta$ itself has no physical meaning because the ``Gibbs state" given by $e^{-\beta \mathcal{L}}/\mathrm{Tr}[e^{-\beta \mathcal{L}}]$ is neither Hermitian nor positive-semidefinite.
It should be considered as a mathematical tool for extracting the large-scale structure of non-steady eigenmodes.

In predicting the qualitative behavior of $\langle \mathcal{O} \rangle_\beta$, it is useful to invoke the analogy with the conventional quantum phase transition in Hermitian systems \cite{Sachdev}.
Figures \ref{Fig-QPT-diagram}(a-1) and (b-1) show the expected phase diagram with respect to the control parameter $g$ and the fictitious temperature $\beta^{-1}$.
The order parameter $\langle \mathcal{O} \rangle_\beta$ takes a nonzero value in the ordered region, but it vanishes in the disordered region.
Figures \ref{Fig-QPT-diagram}(a-2) and (b-2) show the Liouvillian spectra for $g < g_\mathrm{c}$.
The red (blue) region represents the eigenvalues corresponding to the ordered (disordered) eigenmode.
As in Hermitian systems, there are two cases, depending on whether the order exists only at ``zero temperature" $\beta=\infty$ (see Fig.~\ref{Fig-QPT-diagram}(a)) or at finite temperature $\beta<\infty$ (see Fig.~\ref{Fig-QPT-diagram}(b)).
In the former case, the phase transition occurs only in the eigenmodes with the smallest real part of eigenvalues, and these modes are denoted as $|\Phi_{\mathrm{mc}}^{\mathrm{R}})$ and $|\Phi_{\mathrm{mc}}^{\mathrm{L}})$.
Then, the zero-temperature limit of Eq.~\eqref{def_canonical_average} reads
\begin{equation}
	\lim_{\beta \to \infty} \langle \mathcal{O} \rangle_\beta = (\Phi_{\mathrm{mc}}^{\mathrm{L}}| \mathcal{O} |\Phi_{\mathrm{mc}}^{\mathrm{R}}).
\end{equation}
Since the eigenmodes $|\Phi_{\mathrm{mc}}^{\mathrm{R}})$ and $|\Phi_{\mathrm{mc}}^{\mathrm{L}})$ have the largest frequency or the highest coherence, we call them as the most coherent modes of the Liouvillian.
If the order exists at finite temperature, other eigenmodes close to the most coherent mode are also ordered (see Fig.~\ref{Fig-QPT-diagram}(b-2)).
Lower-dimensional systems are expected to correspond to the case (a) and higher-dimensional systems to the case (b).
The dashed lines in Figs.~\ref{Fig-QPT-diagram}(a-1) and (b-1) represent the boundaries of the ``quantum critical region", which is characterized by the absence of quasiparticle-like excitations \cite{Sachdev}.
However, the physical meaning of the quantum critical region in this case is unclear at present, because it includes non-steady eigenmodes far from the steady state.

\section{Model}
\label{sec:model}

We introduce a prototypical model that exhibits the phase transition of non-steady eigenmodes, the transverse field Ising model under dephasing.
The Hamiltonian is given by
\begin{equation}
	H = -J \sum_{\langle j k \rangle} \sigma^z_j \sigma^z_k - g \sum_{j} \sigma^x_j,
\end{equation}
where $\sigma^{\mu}_j \ (\mu=x,y,z)$ denote the Pauli matrices at site $j$ and $\langle j k \rangle$ represents a pair of nearest-neighbor sites.
$J \ (>0)$ and $g$ represent the strength of the exchange interaction and the transverse field, respectively.
Suppose that each spin is affected by dephasing with a rate $\gamma$.
The corresponding jump operator at site $j$ is given by
\begin{equation}
	L_j = \sqrt{\gamma} \sigma^z_j.
	\label{jump_operator_dephasing}
\end{equation}
The Liouvillian is written in the vector representation as
\begin{align}
	\mathcal{L} &= -J \sum_{\langle jk \rangle} (\sigma^z_{j,+} \sigma^z_{k,+} - \sigma^z_{j,-} \sigma^z_{k,-}) - g \sum_{j} (\sigma^x_{j,+} - \sigma^x_{j,-}) \nonumber \\
	&\quad + i \gamma \sum_j \sigma^z_{j,+} \sigma^z_{j,-} - i \gamma N,
	\label{Liouvillian_Ising}
\end{align}
where $\sigma^{\mu}_{j,+(-)}$ acts on the first (second) Hilbert space of $\mathcal{H} \otimes \mathcal{H}$, and $N$ represents the number of spins.

The transverse field Ising model under dephasing is equivalent to the usual Ising model affected by a fluctuating longitudinal field,
\begin{equation}
	\mathcal{H}(t) = -J \sum_{\langle j k \rangle} \sigma^z_j \sigma^z_k - g \sum_{j} \sigma^x_j + \sqrt{\gamma} \sum_{j} \xi_j(t) \sigma^z_j,
	\label{H_Ising_fluc_field}
\end{equation}
where $\xi_j(t)$ represent Gaussian white-noise processes with $\langle \langle \xi_j(t) \rangle \rangle=0$ and $\langle \langle \xi_j(t)\xi_k(t') \rangle \rangle = \delta_{jk} \delta(t-t')$.
Here, $\langle \langle ... \rangle \rangle$ denotes the average with respect to the noise $\xi_j(t)$.
The time evolution of the state vector $\ket{\psi(t)}$ reads $i\partial_t \ket{\psi(t)} = \mathcal{H}(t) \ket{\psi(t)}$.
The density matrix $\rho(t)$ can be obtained by averaging over the noise, $\rho(t) = \langle \langle \ket{\psi(t)}\bra{\psi(t)} \rangle \rangle$.
It can be shown that the time evolution of $\rho(t)$ is given by the GKSL master equation with the jump operator \eqref{jump_operator_dephasing} (see Refs.~\cite{Pichler-13, Stannigel-14, Cai-13} for details of the derivation).
This type of master equation can also be derived, under certain conditions, on the assumption that the environment is a set of independent harmonic oscillators and that the system-environment coupling is linear with respect to the bosonic annihilation and creation operators \cite{Ban-10}.
Experimental realization of the transverse field Ising model is possible with trapped ions \cite{Lanyon-09} and Rydberg atoms \cite{Schauss-12, Sanchez-18, Lienhard-18, Borish-20}, where fluctuations in the trapping field lead to dephasing of the qubit.
Note that for the GKSL master equation to be valid, the correlation time $\tau_\mathrm{c}$ of the noise $\xi_j(t)$ must be much shorter than the time scales of the system.
Since the time scales of the system are characterized by $1/J$, $1/g$, and $1/\gamma$, the legitimate region of the parameters is given by $J, g, \gamma \ll \tau_\mathrm{c}^{-1}$.

\begin{figure}
	\centering
	\includegraphics[width=0.45\textwidth]{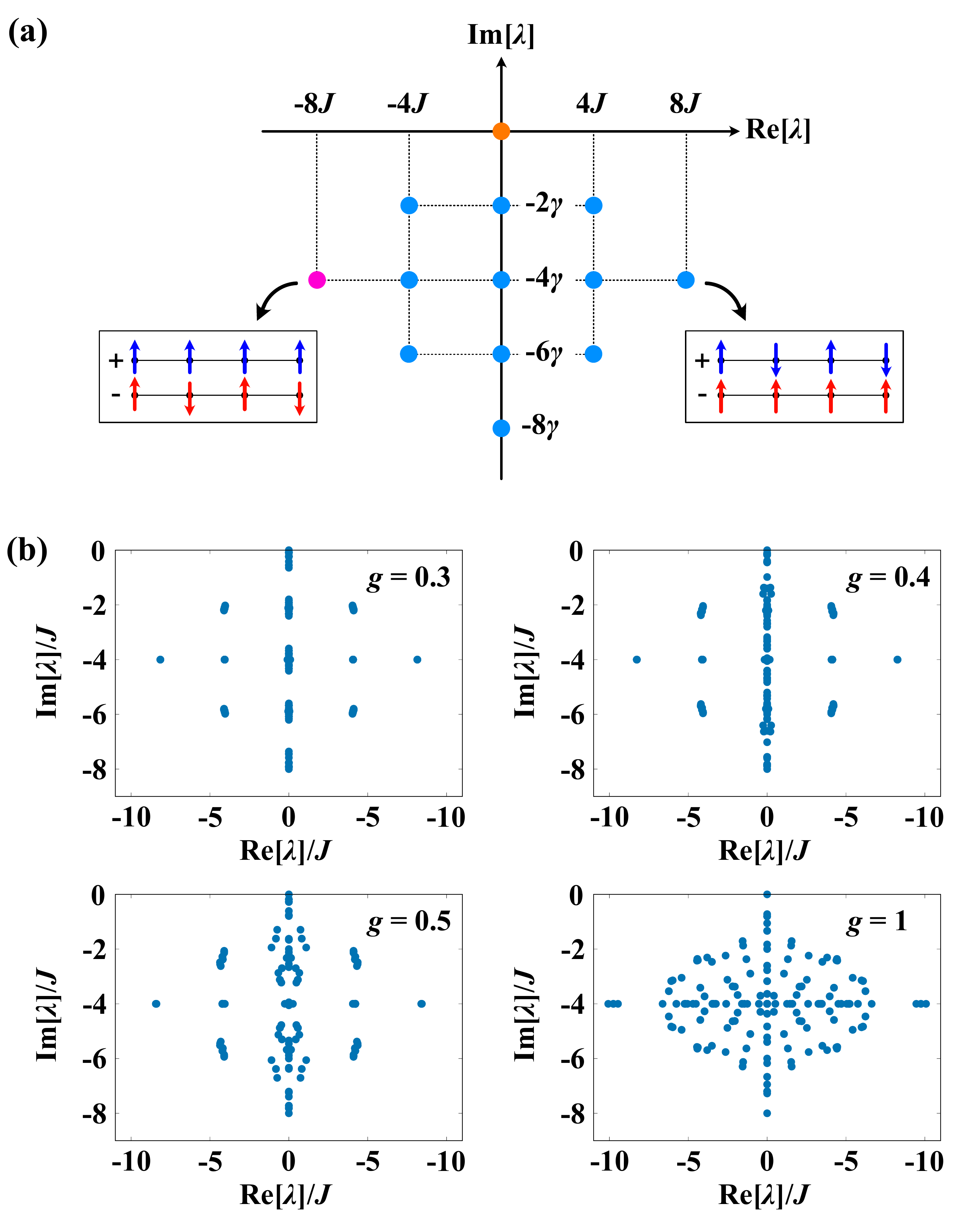}
	\caption{Liouvillian spectra of the transverse field Ising chain under dephasing with $N=4$.
		(a) Spectrum for $g=0$.
		The left-most red point corresponds to the most coherent mode. 
		The insets show the spin configuration of the eigenmodes with the largest and smallest real part of eigenvalues.
		(b) Spectra for $g=0.3$, $0.4$, $0.5$, and $1$ with $J=\gamma=1$.}
	\label{Fig-spec-Ising}
\end{figure}

In the absence of the transverse field $(g=0)$, the most coherent modes of $\mathcal{L}$ are the product states of the ferromagnetic state and N\'eel state, which are four-fold degenerate due to the $\mathbb{Z}_2$ symmetry.
In particular, for the one-dimensional case with the periodic boundary condition, the most coherent modes are written as
\begin{align}
	|\Phi_{\mathrm{mc}}^{\mathrm{R}}) &= \ket{\up \up \up \cdots \up} \otimes \ket{\up \down \up \cdots \down}, \nonumber \\ 
	&\quad \ket{\up \up \up \cdots \up} \otimes \ket{\down \up \down \cdots \up}, \nonumber \\ 
	&\quad \ket{\down \down \down \cdots \down} \otimes \ket{\up \down \up \cdots \down}, \nonumber \\ 
	&\quad \ket{\down \down \down \cdots \down} \otimes \ket{\down \up \down \cdots \up},
	\label{MCM_Ising_zero_g}
\end{align}
where $N$ is assumed to be even.
The corresponding eigenvalue is given by $\lambda = -2JN - i \gamma N$.
In the presence of infinitesimal $g$, the four-fold degeneracy of Eq.~\eqref{MCM_Ising_zero_g} is lifted, and $|\Phi_{\mathrm{mc}}^{\mathrm{R}})$ is the superposition of each term in Eq.~\eqref{MCM_Ising_zero_g} with equal weight.
Figure \ref{Fig-spec-Ising}(a) shows the Liouvillian spectrum for the one-dimensional case with $g=0$ and $N=4$.
Note that each eigenvalue is highly degenerate.
For example, the zero eigenvalue is $2^N$-fold degenerate because the corresponding eigenmodes are the product states of two copies of the same spin configuration.
As $g$ increases, each cluster of eigenvalues becomes elongated, and eventually they merge into a large cluster (see Fig.~\ref{Fig-spec-Ising}(b)).
It should also be noted that the spectrum has the dihedral symmetry with respect to the vertical line $\mathrm{Re}[\lambda] = 0$ and the horizontal line $\mathrm{Im}[\lambda] = -\gamma N$ due to the PT symmetry of the Liouvillian \cite{Prosen-12}.

Conversely, for $g \gg J, \gamma$, the most coherent mode is the paramagnetic state in which all spins are parallel or antiparallel in the $x$-direction:
\begin{equation}
	|\Phi_{\mathrm{mc}}^{\mathrm{R}}) \simeq \ket{\rightarrow \rightarrow \cdots \rightarrow} \otimes \ket{\leftarrow \leftarrow \cdots \leftarrow},
	\label{MCM_Ising_zero_J}
\end{equation}
where $\ket{\rightarrow} = (\ket{\up} + \ket{\down})/\sqrt{2}$ and $\ket{\leftarrow} = (\ket{\up} - \ket{\down})/\sqrt{2}$.
To distinguish between the ferromagnetic and paramagnetic states, we define the magnetization for $+$ spins by
\begin{equation}
	M_+^z := \frac{1}{N} \sum_{j=1}^N \sigma^z_{j,+}.
	\label{def_M}
\end{equation}
Here, note that the expectation value of $M_+^z$ vanishes due to the $\mathbb{Z}_2$ symmetry, $(\Phi_{\mathrm{mc}}^{\mathrm{L}}| M_+^z |\Phi_{\mathrm{mc}}^{\mathrm{R}}) = 0$.
Instead, we consider the squared magnetization,
\begin{equation}
\mathcal{O} := (M_+^z)^2.
\end{equation}
If the most coherent mode $|\Phi_{\mathrm{mc}}^{\mathrm{R}})$ is the paramagnetic state given by Eq.~\eqref{MCM_Ising_zero_J}, the spin correlation function decays exponentially, 
\begin{equation}
(\Phi_{\mathrm{mc}}^{\mathrm{L}}| \sigma^z_{j,+} \sigma^z_{k,+} |\Phi_{\mathrm{mc}}^{\mathrm{R}}) \sim e^{-d_{jk}/\xi}, 
\end{equation}
where $d_{jk}$ is the distance between sites $j$ and $k$, and $\xi$ is the correlation length.
On the other hand, if the most coherent mode $|\Phi_{\mathrm{mc}}^{\mathrm{R}})$ is the ferromagnetic state given by Eq.~\eqref{MCM_Ising_zero_g}, the spin correlation function converges to a nonzero value,
\begin{equation}
\lim_{d_{jk} \to \infty} (\Phi_{\mathrm{mc}}^{\mathrm{L}}| \sigma^z_{j,+} \sigma^z_{k,+} |\Phi_{\mathrm{mc}}^{\mathrm{R}}) = m^2 \neq 0.
\end{equation}
Thus, there exists a critical value $g_\mathrm{c}$ such that $(\Phi_{\mathrm{mc}}^{\mathrm{L}}|\mathcal{O}|\Phi_{\mathrm{mc}}^{\mathrm{R}}) = m^2 \neq 0$ for $g<g_\mathrm{c}$ and $(\Phi_{\mathrm{mc}}^{\mathrm{L}}|\mathcal{O}|\Phi_{\mathrm{mc}}^{\mathrm{R}}) = 0$ for $g>g_\mathrm{c}$ in the thermodynamic limit $N \to \infty$.
In the following, we confirm the existence of the phase transition in the most coherent mode using mean-field analysis and finite-size numerics.

\section{Mean-field analysis}
\label{sec:mean_field}

\begin{figure*}
	\centering
	\includegraphics[width=\textwidth]{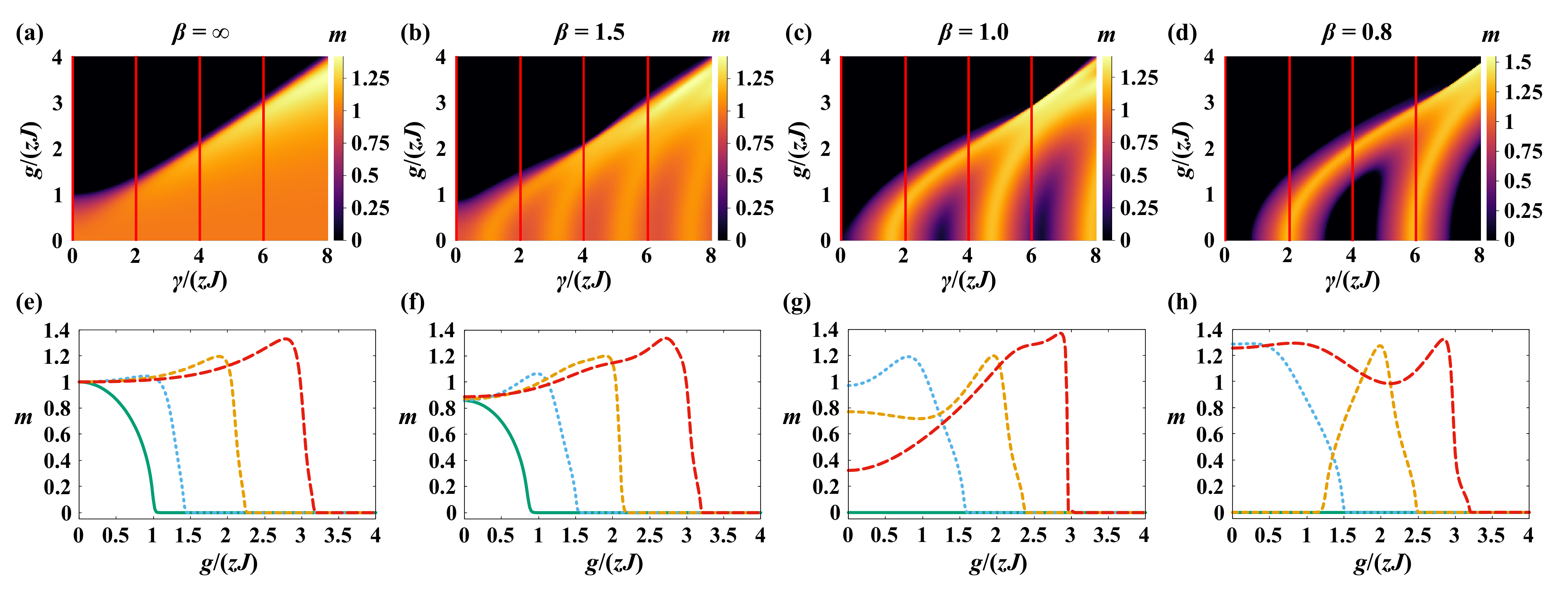}
	\caption{Magnetization $m := (m_\mathrm{A} + m_\mathrm{B})/2 = \mathrm{Re}[m_\mathrm{A}]=\mathrm{Re}[m_\mathrm{B}]$ calculated by the mean-field approximation with $zJ=1$ and $\beta=\infty$, $1.5$, $1.0$, and $0.8$.
		(a)--(d) Magnetization as a function of the dephasing rate $\gamma$ and the transverse field $g$.
		The bright region corresponds to the ferromagnetic phase.
		(e)--(h) Magnetization as a function of $g$ for $\gamma=0$ (solid line: green), $\gamma=2$ (dotted line: blue), $\gamma=4$ (short dashed line: orange), and $\gamma=6$ (long dashed line: red).
		The vertical red lines in (a)--(d) represent the lines of $\gamma=0$, $2$, $4$, and $6$.}
	\label{Fig-MF-phase-diagram}
\end{figure*}

\begin{figure*}
	\centering
	\includegraphics[width=\textwidth]{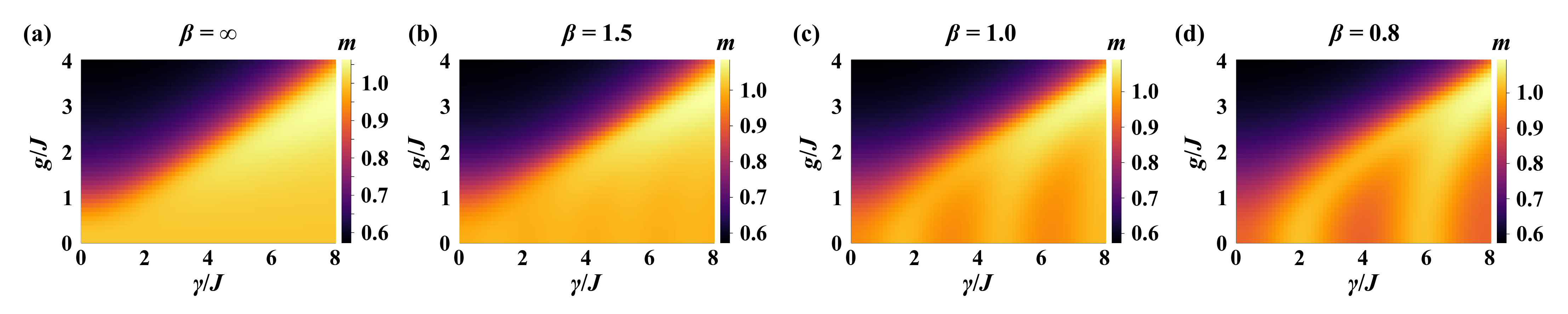}
	\caption{Magnetization $m$ of the one-dimensional dissipative Ising model as a function of the dephasing rate $\gamma$ and the transverse field $g$. 
		The parameters are $J=1$, and $\beta=\infty$, $1.5$, $1.0$, and $0.8$.
		The spin number is $N=4$.
		A clear correlation can be seen with the phase diagram based on the mean-field approximation in Figs.~\ref{Fig-MF-phase-diagram}(a)--(d).}
	\label{Fig-1D-phase-diagram}
\end{figure*}

We apply the mean-field approximation to the non-steady eigenmodes of the dissipative Ising model.
Here, we consider a hypercubic lattice of arbitrary dimension. 
To describe the N\'eel order, the lattice points are divided into the A-sublattice and the B-sublattice.
It is convenient to introduce the following unitary transformation:
\begin{equation}
	U := \prod_j \exp\left( i\frac{\pi}{2} \sigma_{j,-}^z \right) \prod_{j \in \mathrm{B}} \exp\left( i\frac{\pi}{2} \sigma_{j,-}^x \right),
	\label{def_U}
\end{equation}
where $\prod_{j \in \mathrm{B}}$ represents the product over the lattice sites belonging to the B-sublattice.
The first term in Eq.~\eqref{def_U} flips the $x$-component of all spins and the second term flips the $z$-component of the B-sublattice spins.
The Liouvillian is transformed as
\begin{align}
	\tilde{\mathcal{L}} := U^{\dag} \mathcal{L} U &= -J \sum_{\langle jk \rangle} (\sigma^z_{j,+} \sigma^z_{k,+} + \sigma^z_{j,-} \sigma^z_{k,-}) \nonumber \\
	&\quad - g \sum_{j} (\sigma^x_{j,+} + \sigma^x_{j,-}) \nonumber \\
	&\quad + i \gamma \sum_j \mathrm{sgn}(j) \sigma^z_{j,+} \sigma^z_{j,-} - i \gamma N,
	\label{Liouvillian_Ising_ferro}
\end{align}
where $\mathrm{sgn}(j)=1$ for the A-sublattice and $\mathrm{sgn}(j)=-1$ for the B-sublattice.
The transformed Liouvillian \eqref{Liouvillian_Ising_ferro} is symmetric with respect to the exchange of $\sigma^\mu_{j,+}$ and $\sigma^\mu_{j,-}$.
Thus, the most coherent mode for $g=0$ is the tensor product of the ferromagnetic states.

Let us define the ``Gibbs state" with a fictitious inverse temperature $\beta$ by
\begin{equation}
	\rho_\beta = \frac{e^{-\beta \tilde{\mathcal{L}}}}{\mathrm{Tr}\left[e^{-\beta \tilde{\mathcal{L}}}\right]}.
\end{equation}
Note that $\rho_\beta$ is not a physical density matrix because it is neither Hermitian nor positive-semidefinite.
The reduced density matrix for lattice site $j$ is defined by $\rho_{\beta, j} = \mathrm{Tr}_{\neq j}[\rho_\beta]$, where $\mathrm{Tr}_{\neq j}$ represents the trace for all spins except site $j$.
In the mean-field analysis, the reduced density matrix is approximated by
\begin{equation}
	\rho_{\beta, j} \simeq \frac{e^{-\beta \tilde{\mathcal{L}}_j^\mathrm{MF}}}{\mathrm{Tr}_j \left[ e^{-\beta \tilde{\mathcal{L}}_j^\mathrm{MF}} \right]},
\end{equation}
with
\begin{align}
	\tilde{\mathcal{L}}_j^\mathrm{MF} &= - J \sum_{\langle k \rangle_j} (\sigma^z_{j,+} m_{k,+} + \sigma^z_{j,-} m_{k,-}) - g (\sigma^x_{j,+} + \sigma^x_{j,-}) \nonumber \\
	&\quad + i \gamma \mathrm{sgn}(j) \sigma^z_{j,+} \sigma^z_{j,-} - i \gamma,
\end{align}
where $\langle k \rangle_j$ represents the nearest neighboring sites around $j$ and $\mathrm{Tr}_j$ represents the trace for spins at site $j$.
The magnetization $m_{k, \pm} := \mathrm{Tr}_k [\sigma^z_{k, \pm} \rho_{\beta, k}]$ at site $k$ is determined self-consistently.
Note that $m_{k, \pm}$ is generally complex-valued.
Since $m_{k, \pm}$ depends only on the sublattice to which $k$ belongs, we denote it as $m_\mathrm{A}$ or $m_\mathrm{B}$.
Then, the mean-field Liouvillian is given by
\begin{align}
	\tilde{\mathcal{L}}_\mathrm{A}^\mathrm{MF} &= - zJ m_\mathrm{B} (\sigma^z_{+} + \sigma^z_{-}) - g (\sigma^x_{+} + \sigma^x_{-}) \nonumber \\
	&\quad + i \gamma \sigma^z_{+} \sigma^z_{-} - i \gamma,
	\label{mf_Liouvillian_A}
\end{align}
\begin{align}
	\tilde{\mathcal{L}}_\mathrm{B}^\mathrm{MF} &= - zJ m_\mathrm{A} (\sigma^z_{+} + \sigma^z_{-}) - g (\sigma^x_{+} + \sigma^x_{-}) \nonumber \\
	&\quad - i \gamma \sigma^z_{+} \sigma^z_{-} - i \gamma,
	\label{mf_Liouvillian_B}
\end{align}
where $z$ is the coordinate number of the lattice and we have omitted the site index of the spin operator.
The self-consistent equations read
\begin{equation}
	m_\mathrm{A} = \frac{\mathrm{Tr}_1 \left[\sigma_+^z e^{-\beta \tilde{\mathcal{L}}_\mathrm{A}^\mathrm{MF}}\right]}{\mathrm{Tr}_1 \left[ e^{-\beta \tilde{\mathcal{L}}_\mathrm{A}^\mathrm{MF}} \right]}, \quad m_\mathrm{B} = \frac{\mathrm{Tr}_1 \left[\sigma_+^z e^{-\beta \tilde{\mathcal{L}}_\mathrm{B}^\mathrm{MF}}\right]}{\mathrm{Tr}_1 \left[ e^{-\beta \tilde{\mathcal{L}}_\mathrm{B}^\mathrm{MF}} \right]},
	\label{mf_self_consistent}
\end{equation}
where ``$\mathrm{Tr}_1$" represents the trace over the one-site Hilbert space with a basis set $\{ \ket{\up} \otimes \ket{\up}, \ket{\up} \otimes \ket{\down}, \ket{\down} \otimes \ket{\up}, \ket{\down} \otimes \ket{\down} \}$.
From Eqs.~\eqref{mf_Liouvillian_A}, \eqref{mf_Liouvillian_B}, and \eqref{mf_self_consistent}, the following relation holds:
\begin{equation}
	m_\mathrm{A} = m_\mathrm{B}^*.
	\label{m_A_m_B_relation}
\end{equation}
The self-consistent equation \eqref{mf_self_consistent} together with Eqs.~\eqref{mf_Liouvillian_A}, \eqref{mf_Liouvillian_B}, and \eqref{m_A_m_B_relation} is numerically solved by iteration.

Figure \ref{Fig-MF-phase-diagram} shows the averaged magnetization $m :=(m_\mathrm{A} + m_\mathrm{B})/2$ as a function of the dephasing rate $\gamma$ and the transverse field $g$.
From Eq.~\eqref{m_A_m_B_relation}, the magnetization $m$ is real.
For Hermitian systems, the magnetization satisfies $-1 \leq m \leq 1$.
On the other hand, note that in this case $m$ becomes greater than $1$ when $\gamma$ and $g$ are large.
This is because the right and left eigenmodes are not equal due to the non-Hermiticity of the Liouvillian.
In Figs.~\ref{Fig-MF-phase-diagram}(a)--(d), the bright region with $m>0$ corresponds to the ferromagnetic (ordered) phase, and the dark region with $m=0$ corresponds to the paramagnetic (disordered) phase.
The phase boundary determines the critical field $g_\mathrm{c}(\gamma)$.
For $\gamma=0$, since the model is identical to the decoupled copies of the transverse field Ising model, it exhibits a quantum phase transition at $g=1$ at zero temperature ($\beta=\infty$).
The critical field $g_\mathrm{c}$ for $\gamma=0$ decreases as the temperature increases, and it eventually vanishes at $\beta=1$.
A remarkable feature of magnetization is its oscillatory behavior as a function of $\gamma$.
In the absence of the transverse field ($g=0$), the magnetization $m(\gamma, g)$ has a periodicity,
\begin{equation}
	m\left(\gamma + \frac{\pi}{\beta}, 0\right) = m(\gamma, 0),
	\label{m_periodicity}
\end{equation}
which directly follows from the self-consistent equation \eqref{mf_self_consistent}.
Since for $\beta<1$, $m(\gamma, g)$ vanishes near $\gamma=g=0$, the periodicity \eqref{m_periodicity} implies the existence of infinitely many islands of paramagnetic phase, as shown in Fig.~\ref{Fig-MF-phase-diagram}(d).
In this case, the phase diagram has a reentrant structure.
For $\beta=0.8$ and $\gamma=4$, the transition from the paramagnetic phase to the ferromagnetic phase occurs at $g\simeq1.1$ with increasing $g$, and the second transition to the paramagnetic phase occurs at $g\simeq2.5$ (see Figs.~\ref{Fig-MF-phase-diagram}(d) and (h)).

\section{Numerical results for 1D model}
\label{sec:numerical_1D}

We study the phase structure of the most coherent mode by the finite-size numerics of the one-dimensional dissipative Ising model.
We assume the periodic boundary condition and an even number of spins.
First, let us consider the phase diagram at finite temperature.
The magnetization $M_+^z$ for $+$ spins is defined by Eq.~\eqref{def_M}, and we consider the average of the squared magnetization,
\begin{equation}
	m^2 := \frac{\mathrm{Tr}\left[(M_+^z)^2 e^{-\beta \tilde{\mathcal{L}}}\right]}{\mathrm{Tr}\left[e^{-\beta \tilde{\mathcal{L}}}\right]}.
	\label{def_m_finite_size}
\end{equation}
The magnetization $m$ is given by the square root of Eq.~\eqref{def_m_finite_size}, which corresponds to $(m_\mathrm{A} + m_\mathrm{B})/2$ of the mean-field analysis in Sec.~\ref{sec:mean_field}, where $m_\mathrm{A} \:(m_\mathrm{B})$ is the magnetization at the sublattice A (B).

Figure \ref{Fig-1D-phase-diagram} shows the magnetization $m$ as a function of the dephasing rate $\gamma$ and the transverse field $g$. 
The spin number is $N=4$.
The clear correlation between Figs.~\ref{Fig-MF-phase-diagram} and \ref{Fig-1D-phase-diagram} confirms the validity of the mean-field approximation.
In particular, one can observe a precursor of the reentrant structure of the phase diagram in Fig.~\ref{Fig-1D-phase-diagram}(d).
As in the case of the conventional transverse field Ising model, no true long-range order is expected to exist at finite temperature in one dimension.
Thus, the magnetization for $\beta < \infty$ should vanish in the thermodynamic limit $N \to \infty$.

\begin{figure}
	\centering
	\includegraphics[width=0.45\textwidth]{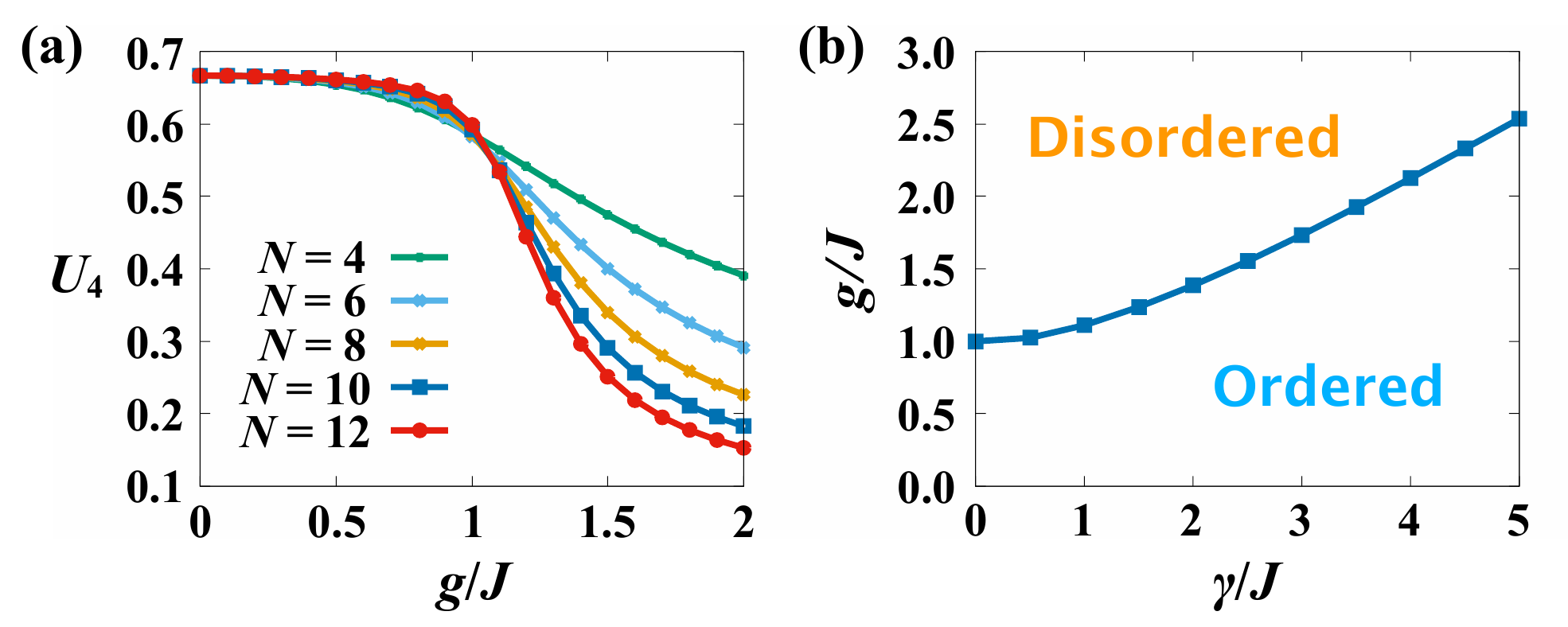}
	\caption{(a) Binder cumulant $U_4$ of the most coherent mode as a function of the transverse field $g$ with $\gamma=1$.
		The systems sizes are $N=4$, $6$, $8$, $10$, and $12$.
		The cross point of $U_4$ for different system sizes indicates the critical field $g_\mathrm{c}$.
		(b) Critical field $g_\mathrm{c}$ as a function of $\gamma$. 
		The regions below and above the curve correspond to the long-range ordered phase and the disordered phase, respectively.}
	\label{Fig-1D-binder}
\end{figure}

Next, we focus on the phase transition of the most coherent mode ($\beta=\infty$).
It is hard to calculate the most coherent mode through the numerical diagonalization of the Liouvillian for $N>8$.
Thus, we consider the imaginary time evolution of the master equation,
\begin{equation}
	\frac{d}{dt}|\rho) = -\tilde{\mathcal{L}}|\rho).
	\label{master_eq_imag_time}
\end{equation}
By integrating Eq.~\eqref{master_eq_imag_time} for a sufficiently long time, $|\rho)$ converges to the most coherent mode $|\Phi_{\mathrm{mc}}^{\mathrm{R}})$.
To determine the critical field $g_\mathrm{c}(\gamma)$, we define the Binder cumulant $U_4$ for the most coherent mode as
\begin{equation}
	U_4 := 1 - \frac{\mathrm{tr}\left[(M^z)^4 \Phi_{\mathrm{mc}}^{\mathrm{R}}\right]}{3\mathrm{tr}\left[(M^z)^2 \Phi_{\mathrm{mc}}^{\mathrm{R}}\right]^2},
	\label{def_binder}
\end{equation}
where $\Phi_{\mathrm{mc}}^{\mathrm{R}}$ is the original matrix representation of $|\Phi_{\mathrm{mc}}^{\mathrm{R}})$ and 
\begin{equation}
M^z = \frac{1}{N} \sum_{j=1}^N \sigma^z_{j}
\end{equation}
is the magnetization for the original spin operator.
Note that the trace ``$\mathrm{tr}$" in Eq.~\eqref{def_binder} is taken over the original Hilbert space $\mathcal{H}$.
In the thermodynamic limit, the Binder cumulant $U_4$ takes the value $2/3$ in the long-range ordered phase and zero in the disordered phase.

Figure \ref{Fig-1D-binder}(a) shows the Binder cumulant $U_4$ as a function of the transverse field $g$ with $\gamma=1$.
For sufficiently large system sizes, it is known that $U_4$ for different system sizes intersect at the critical field $g_\mathrm{c}$.
To determine $g_\mathrm{c}$, let us denote the cross point of $U_4$ for two system sizes $N$ and $N-2$ as $g_\mathrm{c}^{(N)}$.
The critical field is given by $g_\mathrm{c}=\lim_{N\to\infty} g_\mathrm{c}^{(N)}$.
We fit $g_\mathrm{c}^{(N)}$ for $N=6$, $8$, $10$, $12$ with the algebraic function $f(N) = a+bN^{-c} \:(c>0)$, where $g_\mathrm{c}=a$.
Figure \ref{Fig-1D-binder}(b) shows $g_\mathrm{c}$ determined by this procedure as a function of the dephasing rate $\gamma$.
The regions of $g<g_\mathrm{c}$ and $g>g_\mathrm{c}$ correspond to the long-range ordered phase and the disordered phase, respectively.
Note that for $\gamma=0$, $g_\mathrm{c}=1$ because the model is equivalent to two copies of the conventional transverse field Ising model.
The curve of $g_\mathrm{c}$ in Fig.~\ref{Fig-1D-binder}(b) coincides with the phase boundary in Fig.~\ref{Fig-1D-phase-diagram}(a) for small system size.
These results provide evidence for the existence of an order-disorder transition of the most coherent mode in one dimension.

\section{Relaxation dynamics of a cat state}
\label{sec:relaxation}

We discuss how the phase transition of the most coherent mode affects the relaxation dynamics of the system.
As an initial state, we consider the following pure state:
\begin{equation}
	\ket{\psi_\mathrm{ini}} = \frac{1}{\sqrt{2}} (\ket{\mathrm{F}} + \ket{\mathrm{N}}),
	\label{cat_state}
\end{equation}
where $\ket{\mathrm{F}}=\ket{\up \up \up \cdots \up}$ is a ferromagnetic state and $\ket{\mathrm{N}}=\ket{\up \down \up \cdots \down}$ is a N\'eel state.
Superpositions of two macroscopically different quantum states, such as Eq.~\eqref{cat_state}, are called the ``Schr\"odinger's cat states" \cite{Leibfried-05, Monz-11, Omran-19, Song-19}, which are an important resource in quantum computing and quantum communication.
The corresponding density matrix reads 
\begin{align}
	\rho_\mathrm{ini} &= \ket{\psi_\mathrm{ini}} \bra{\psi_\mathrm{ini}} \nonumber \\ 
	&= \frac{1}{2} (\ket{\mathrm{F}} \bra{\mathrm{F}} + \ket{\mathrm{N}} \bra{\mathrm{N}} + \ket{\mathrm{F}} \bra{\mathrm{N}} + \ket{\mathrm{N}} \bra{\mathrm{F}}).
	\label{rho_ini}
\end{align}
In the limit of $g \to 0$, the most coherent mode is the equal superposition of Eq.~\eqref{MCM_Ising_zero_g}.
Thus, the overlap between the initial state $\rho_\mathrm{ini}$ and the most coherent mode $\Phi_{\mathrm{mc}}^{\mathrm{R}}$ is given by
\begin{equation}
	(\rho_\mathrm{ini}| \Phi_{\mathrm{mc}}^{\mathrm{R}}) = \frac{1}{4}.
\end{equation}
From the eigenmode expansion \eqref{rho_expansion}, the large overlap between the initial state and the most coherent mode means that the time evolution of the density matrix is dominated by the most coherent mode.
Thus, the density matrix is expected to exhibit coherent oscillations at the frequency of the most coherent mode.
In the opposite case $g \gg J$, since the most coherent mode is approximately given by Eq.~\eqref{MCM_Ising_zero_J}, the overlap with the initial state is significantly small.
Thus, the density matrix is expected to exhibit monotonic relaxation.
In summary, when $g$ exceeds the critical field $g_\mathrm{c}$, a crossover from coherent relaxation with oscillations to incoherent relaxation without oscillations is expected to occur.

\begin{figure}
	\centering
	\includegraphics[width=0.45\textwidth]{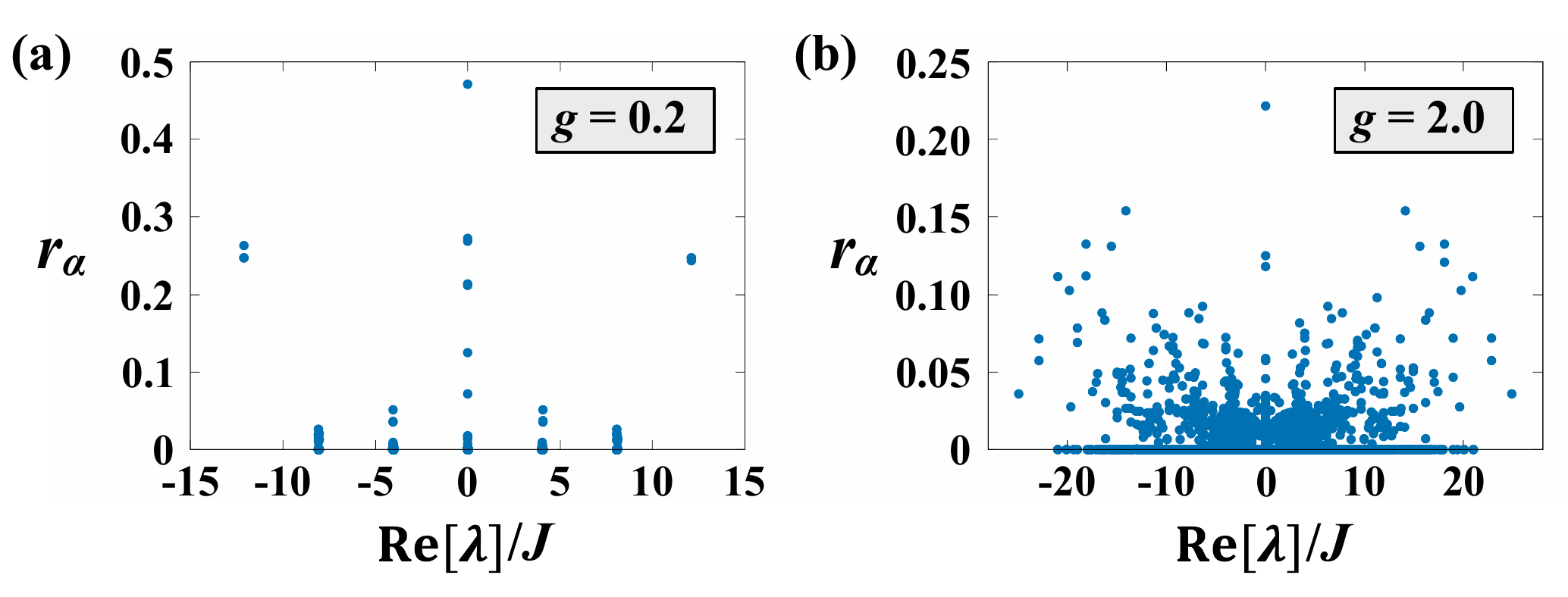}
	\caption{Overlap $r_\alpha = |(\rho_\mathrm{ini}|\Phi_\alpha^{\mathrm{R}})|$ between eigenmodes $|\Phi_\alpha^{\mathrm{R}})$ and the initial state $|\rho_\mathrm{ini})$ for (a) $g=0.2$ and (b) $g=2$.
		The spin number is $N=6$ and the other parameters are $J=\gamma=1$.
		The horizontal axis represents the real part of the eigenvalues $\lambda_\alpha$.
		For a small $g$, the overlap is localized to eigenmodes where the real part of the corresponding eigenvalue is minimum, maximum, or zero.
		On the other hand, for a large $g$, the overlap is delocalized over all eigenmodes.
		Note that $r_\alpha$ is not perfectly symmetric with respect to $\mathrm{Re}[\lambda]=0$.
		This is due to the ambiguity in the choice of eigenmodes in eigenspace when degeneracy occurs.}
	\label{Fig-overlap}
\end{figure}

Figure \ref{Fig-overlap} shows the overlap 
\begin{equation}
	r_\alpha := |(\rho_\mathrm{ini}|\Phi_\alpha^{\mathrm{R}})|
\end{equation}
between eigenmodes $|\Phi_\alpha^{\mathrm{R}})$ and the initial state $|\rho_\mathrm{ini})$ given by Eq.~\eqref{rho_ini}.
For $g \ll J$, the overlap has large values for eigenmodes with $\mathrm{Re}[\lambda_\alpha]=0, \pm 2NJ$ (see Fig.~\ref{Fig-overlap}(a)), because the eigenmodes with $\mathrm{Re}[\lambda_\alpha]=0$, $-2NJ$, and $2NJ$ have large overlaps with $\ket{\mathrm{F}} \bra{\mathrm{F}} + \ket{\mathrm{N}} \bra{\mathrm{N}}$, $\ket{\mathrm{F}} \bra{\mathrm{N}}$, and $\ket{\mathrm{N}} \bra{\mathrm{F}}$, respectively.
This implies that the time evolution of the density matrix starting from $\rho_{\mathrm{ini}}$ is governed by a small number of eigenmodes with frequency $2NJ$.
When $g$ is comparable to $J$ (see Fig.~\ref{Fig-overlap}(b)), the overlap is delocalized over all eigenmodes.
In this case, since the time evolution of the density matrix is given by the superposition of a large number of eigenmodes with various frequencies, an incoherent relaxation without oscillations is expected.

\begin{figure}
	\centering
	\includegraphics[width=0.45\textwidth]{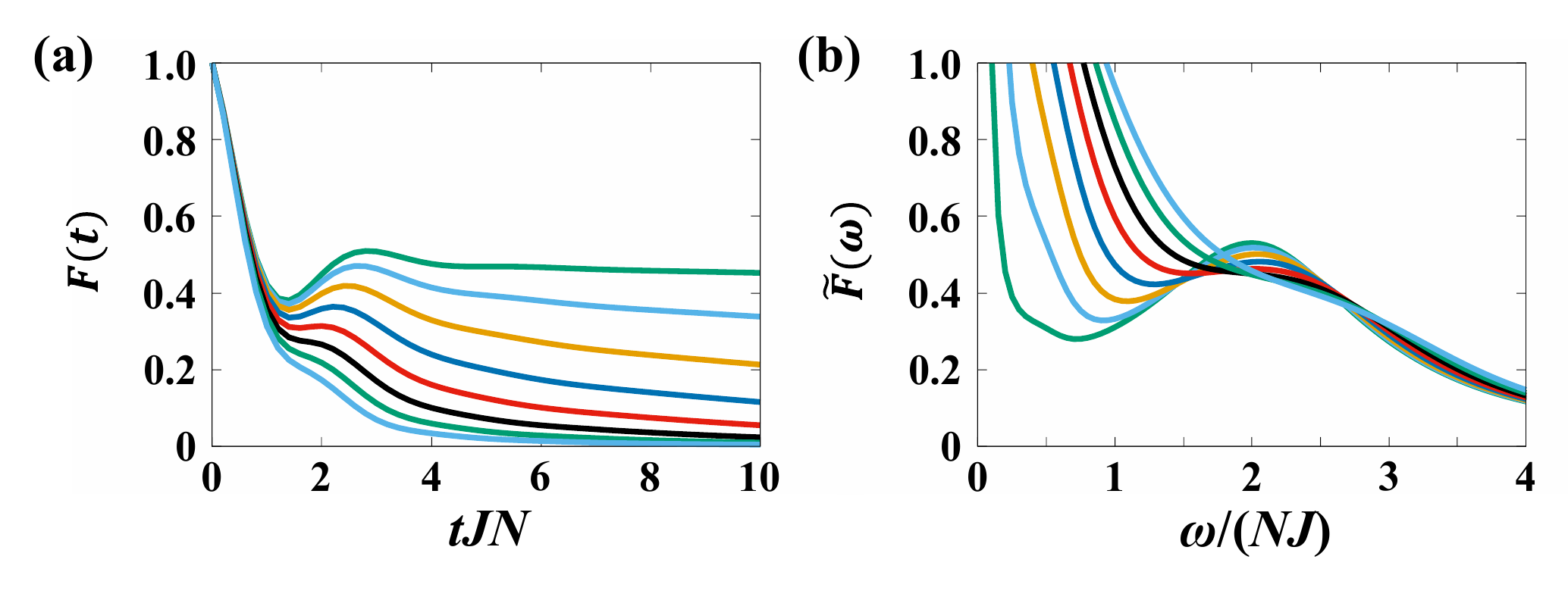}
	\caption{(a) Time evolution of the fidelity $F(t)$ for $g=0.2$, $0.4$, $0.6$, $0.8$, $1.0$, $1.2$, $1.4$, and $1.6$ from top to bottom.
		The spin number is $N=8$ and the other parameters are $J=\gamma=1$.
		(b) Fourier transform of $\tilde{F}(\omega)$ for $g=0.2$, $0.4$, $0.6$, $0.8$, $1.0$, $1.2$, $1.4$, and $1.6$ from bottom to top at $\omega/(NJ)=1$.
		The peak at $\omega=2NJ$ disappears at $g\simeq1.1$, which is close to the critical field $g_\mathrm{c}\simeq1.11$ at which the phase transition of the most coherent mode occurs.}
	\label{Fig-fidelity}
\end{figure}

We consider the fidelity
\begin{equation}
	F(t) := \left( \mathrm{tr}\left[ \left(\rho_\mathrm{ini}^{1/2} \ \rho_t \ \rho_\mathrm{ini}^{1/2} \right)^{1/2} \right] \right)^2,
\end{equation}
which represents the distance between the initial state and the state at time $t$ \cite{Jozsa-94, Liang-19}.
We have $F(0)=1$ at $t=0$ and $F(t) \to D^{-1}$ in the long-time limit $t \to \infty$, where $D=2^N$ is the dimension of the Hilbert space.
Figure \ref{Fig-fidelity}(a) shows the time evolution of the fidelity $F(t)$ for different values of $g$.
Note that the horizontal axis is the rescaled time $tJN$.
We confirm that $F(t)$ with respect to the rescaled time $tJN$ is almost independent of the system size $N$.
For $g<1$, $F(t)$ exhibits temporal oscillations, whereas for $g>1$, $F(t)$ decays monotonically.
To highlight this crossover, in Fig.~\ref{Fig-fidelity}(b), we show the Fourier transform of $F(t)$,
\begin{equation}
	\tilde{F}(\omega) = \int_{-\infty}^{\infty} dt e^{-i\omega t} F(|t|).
\end{equation}
The horizontal axis is the rescaled frequency $\omega/(NJ)$.
For $g<1$, $\tilde{F}(\omega)$ shows a peak at $\omega/(NJ)=2$, which is consistent with the fact that the time evolution is dominated by the most coherent mode with frequency $2NJ$.
As $g$ increases, the peak is smeared and eventually disappears at $g \simeq 1.1$.
The value of $g$ at which the peak of $\tilde{F}(\omega)$ disappears is close to the critical field $g_\mathrm{c}\simeq 1.11$ of the most coherent mode.
These results suggest that the transition of the most coherent mode leads to a qualitative change in the decoherence of a cat state.

\section{Conclusions}
\label{sec:conclusion}

In this study, we found that non-steady eigenmodes of a simple open quantum many-body system exhibit spontaneous symmetry breaking.
The transition from the disordered phase to the symmetry-broken ordered phase occurs only in the vicinity of the most coherent mode, which is the eigenmode with the highest frequency.
This means that the transition does not change the long-time behavior of the system, but only affects the transient relaxation dynamics.
In particular, we demonstrated the crossover from under-damped to over-damped relaxation of a Schr\"odinger's cat state.

We discuss the experimental feasibility of the setup postulated in this study.
The transverse field Ising model can be realized by trapped ions \cite{Lanyon-09} or Rydberg atoms \cite{Schauss-12, Sanchez-18, Lienhard-18, Borish-20}.
The dephasing of the system is caused by different types of noise, such as fluctuations in the magnetic field of the ion-trap device.
Recent advances in qubit manipulation techniques have made it possible to generate Schr\"odinger's cat states with dozens of qubits \cite{Leibfried-05, Monz-11, Omran-19, Song-19}.
The off-diagonal elements of the density matrix, which characterize the coherence of the state, can be obtained by observing the parity oscillations.
We expect that there exists a critical value of the transverse field at which the way the coherence decays changes qualitatively.

The introduction of the Liouvillian canonical ensemble defined by Eq.~\eqref{def_canonical_average} allows various field theoretical approaches, such as mapping to a classical model with an additional dimension, diagrammatic techniques, and renormalization group analysis, which have been developed in the study of conventional quantum phase transitions \cite{Sachdev}.
Such an analysis can be useful in clarifying ``low-energy" excitations near the most coherent mode and the nature of the quantum critical regime shown in Fig.~\ref{Fig-QPT-diagram}.
The above field theoretical approach should not be confused with the Keldysh formalism \cite{Torre-13, Maghrebi-16, Sieberer-16, Young-20}, which is developed to study steady states.
To establish a field theory for non-steady eigenmodes can open a new research direction on the nonequilibrium dynamics of open quantum many-body systems.

\begin{acknowledgments}
	This work was supported by JSPS KAKENHI Grant Number JP22K13983.
\end{acknowledgments}

\end{document}